\documentstyle[12pt,aasms]{article}
\def\etal{{\it et al. }}
\def\kms{km~s$^{-1}$ }
\received{7 June 1993}
\accepted{16 July 1993}
\journalid{415}{1 October 1993}
\articleid{L95}{L98}

\slugcomment{Accepted for publication in {\it The Astrophysical Journal Letters}}

\begin{document}

\title{THE ROTATION CURVES OF GALAXIES AT INTERMEDIATE
REDSHIFT\altaffilmark{1}}

\author{Nicole P. Vogt and Terry Herter}
\affil{Center for Radiophysics and Space Research, Cornell University,
Ithaca, NY 14853}

\author{Martha P. Haynes}
\affil{Center for Radiophysics and Space Research, Cornell University
and National Astronomy and Ionosphere Center\altaffilmark{2},
Ithaca, NY 14853}

\and

\author{St\'ephane Courteau}
\affil{Center for Radiophysics and Space Research, Cornell University,
Ithaca, NY 14853}

\altaffiltext{1}{Based on observations made at the Palomar Observatory as
part of a continuing collaborative agreement between the California
Institute of Technology and Cornell University.} 
\altaffiltext{2}{The National Astronomy and Ionosphere Center is
operated by Cornell University under a management agreement with the
National Science Foundation.}

\hsize 6.2 truein
\begin{abstract} 

We have undertaken a pilot project to measure the rotation velocities
of spiral galaxies in the redshift range 0.18 $\le z \le$ 0.4 using
high dispersion long slit spectroscopy obtained with the Palomar 5m
telescope. One field galaxy and three cluster objects known to have
strong emission lines were observed over wavelength ranges covering the
redshifted lines of [OII], CaII K, H$\beta$, and [OIII]. Two of the
objects show extended line emission that allows the tracing of the
rotation curve in one or more lines.  A line width similar to that
obtained with single dish telescopes for the 21--cm HI line observed in
lower redshift galaxies can be derived from the observed H$\beta$,
[OII], and [OIII] emission by measuring a characteristic width from the
velocity histogram.  These moderately distant galaxies have much
stronger emission lines than typical low--redshift spirals but they
appear to be kinematically similar.  Application of the
Tully-Fisher relation suggests that the two galaxies with rotation
curves are intrinsically brighter at R-band than
nearby galaxies.
\end{abstract}


\section{Introduction}

A major goal of cosmology is to determine the temporal history and fate of
the universe.  Galaxies provide the primary means to accomplish this; thus it
is critically important to understand their formation and evolution.  Between
the present time and the epoch corresponding to a redshift $z \simeq$ 0.4,
significant evolution of the cluster population results in a higher fraction
of blue galaxies at the earlier epoch. Most recent high resolution imaging of
the blue cluster members confirms their spiral nature (Lavery \etal 1992;
Dressler \etal 1993). A number of authors conclude that galaxy--galaxy
interactions play a primary role in producing an enhancement in the blue
galaxy fraction in distant clusters (Thompson 1988; Lavery and Henry 1988;
Lavery \etal 1992) while others suggest that the excess activity in blue
galaxies is associated with the interaction between hot intracluster gas and
infalling galaxies (Dressler \etal 1985; Bothun and Dressler 1986). The
detailed study of the kinematics of the blue cluster galaxies and their
counterparts in the field at similar redshifts will place constraints on both
the formation and evolutionary processes.  

In recent years, the HI Tully--Fisher (TF) relation and its optical analog
have been used by a growing community in attempts to map out the local flow
field.  These peculiar motions hinder determination of the Hubble constant
from nearby galaxies.  However, at distances larger than $z \simeq$ 0.04,
peculiar motions should be negligible with respect to the overall Hubble flow
and the main source of error is the intrinsic scatter in the TF relation
(about 0.3 mag) and the local calibration.  Thus, if the TF
method can be properly understood and applied, it may permit the
determination of redshift--independent distances and evaluation of the Hubble
constant, at intermediate redshift.

The prospects for high redshift application of the TF method using the
H$\alpha$ line have been discussed by van der Kruit and Pickles
(1988).  Detection of the H$\alpha$ line as the distance to the
emitting object increases is hampered not only by the radial fall-off in
emissivity but also because the redshifting of the line from its rest
wavelength places the H$\alpha$ line in the
night--sky--contaminated red portion of the spectrum until it finally
moves out of the optical window at $z \ge$ 0.4.  An alternative
approach, discussed in the context of star formation indicators by
Kennicutt (1992a,b), relies on lines whose rest wavelengths are
significantly lower than that of H$\alpha$, specifically, the emission
lines of [OII], H$\beta$, and [OIII]. All three lines are often (though
{\it not always}) seen in the spectra of nearby spiral galaxies, the same
objects in which H$\alpha$ is detected.

In this {\it Letter}, we report the results of a 
pilot project to conduct high resolution, high--sensitivity 
long--slit spectroscopic observations of galaxies in the
redshift range 0.18 $\le z \le$ 0.4.  In \S \ 2 we discuss our
observational strategy.  A summary of the results, including the
extended rotation curves of two galaxies with $z \simeq$ 0.2, is
presented in \S \ 3. Finally, we discuss the degree of success of
our experiment and the potential for using this technique to derive
an estimate of the Hubble constant out to such distances.

\section{Observations}

Since our objective was to ensure the greatest likelihood for measuring
rotation curves we chose objects in the redshift range 0.2 -- 0.4 known to
have strong emission lines and displaying an optical appearance suggestive of
spiral morphology.  Final observations were constrained by the specific time
available for observation which was primarily determined by weather
conditions.

Long-slit spectra were taken with the Hale 5m telescope using the
Double Spectrograph (Oke and Gunn 1982) in November 1992 and January
1993. Simultaneous blue (3100 -- 5200 \AA) and red (5200 -- 11000 \AA) 
spectra were obtained with 1200 l/mm gratings and the Palomar D52
dichroic beam splitter. A 1\arcsec \ slit was used for the
observations.  The blue and red camera have a spatial scale of
0.78\arcsec/pixel and 0.58\arcsec/pixel, and a dispersion of 0.56
\AA/pixel and 0.82 \AA/pixel respectively. We observed the resolved
[OII] doublet ($\lambda$3726, $\lambda$3729) in the blue camera,
and the [OIII] line pair ($\lambda$4959, $\lambda$5007), and
H$\beta$ ($\lambda$4861) with the red camera. 

An accurate determination of source coordinates proved to be
essential to the success of this project.  Coordinates for
surrounding stars were taken from the HST Guide Star Catalog and
applied exactingly to create a transformation matrix for images of
the clusters into right ascension and declination.  We obtained very
accurate relative coordinates between our target sources, the other
galaxies in the cluster, and the neighboring guide stars.  

The $\sim$21st magnitude sources were completely invisible to the guider
camera when in the slit, and by mischance no stars fell within the fields of
view on our exposures.  During both observing runs weather conditions were
somewhat variable, and the seeing conditions ranged from sub--arcsecond to
greater than 1.5\arcsec.  Under poor seeing conditions ($>$ 1\arcsec) we
repeatedly halted the exposures and returned to the guide star to correct for
possible drift, every ten to twenty-five minutes depending upon the wind
conditions and the hour angle of the telescope.  When the seeing was
sub--arcsecond and it was completely clear, we were able to guide readily on
companion galaxies within the cluster.  To achieve adequate signal-to-noise
in the extended emission (based on experiences observing lower redshift
galaxies) exposure times of two hours were used, weather permitting.  These
times and a rough estimate of the seeing quality for the four distant
galaxies for which we obtained adequate data are included with other relevant
information in Table 1.

\vfill\eject
\section{Results}

We performed the data reduction within the IRAF programming environment,
using standard packages for the initial stages and then linked FORTRAN
programs to fit a Gaussian profile to the line at each spatial position. The
[OII] doublet and [OIII] line pair were each fit as a double Gaussian
profile.  The error bars on the rotation curves displayed in figures 1 and 3
reflect the quality of the Gaussian fits to the data;  we believe them to be
larger than any systematic errors.  The velocity scales on the y-axis measure
the rotational velocity of the galaxy about an estimated central point {\it
in the rest frame of the galaxy}.  A characteristic velocity width, W, is
derived from a histogram of the velocity distribution and is the full width
measured across the inner 80$\%$ of the histogram area. No intensity
weighting is performed. This width is consistent with those estimated from
fitting the maximum of the rotation curve and from visual inspection of the
rotation curves.  Since each of the current observations is distinct in terms
of both intrinsic nature and degree of success, we discuss the results
separately.

\vskip 5pt \noindent {\it Galaxy 2545.3 in SA 68: \ }
The first object is a strong emission line galaxy from the SA 68
field sample of Koo (1985). The particular object designated 2545.3
($z$ = 0.211) was chosen because it has very strong emission of [OII],
H$\beta$, and [OIII]. The morphology of the object is difficult
to establish on a photographic plate made available by Koo and Ellman
but it appears elongated with a small bulge. It is clearly unusual 
given the strength of its emission lines. 

Figure 1 shows the observed variation in velocity as derived for the [OII]
and [OIII] separately. As a demonstration of the data quality, Figure 2
presents the spectral image of the resolved [OII] doublet. An unexpected
companion source was detected 2\arcsec \ away on the sky, an order of
magnitude weaker with no distinguishing features.  The full optical extent of
the galaxy can be seen in [OII], H$\beta$, and [OIII], though the
red--shifted H$\beta$ line falls directly upon a broad OH night sky line from
which it proved to be inseparable.  All four oxygen lines are quite strong.
As seen in Figure 1, the rotation curve is quite symmetric, and both lines
give consistent results. The velocity widths determined independently from
each set of lines agree to within 10 \kms, giving a final value of 115 $\pm
5$ \kms for W.  The inclination of the galaxy, determined from digitized 4m
photographic plate material is 44$^{\circ} \pm 3^{\circ}$, yielding a
corrected width $W_c$ of 166 $\pm 13$ \kms.

\vskip 5pt 
\noindent {\it Galaxies No. 57 and No. 208 in CL 0949+44: \ } 
We detected two galaxies in the cluster CL0949+44 at $z$ = 0.380,
selected for the presence of strong emission from the spectroscopic
catalog of Dressler and Gunn (1992). As discussed by Dressler and Gunn,
this cluster actually contains two separate velocity components and is
populated by the bluest galaxies in their cluster sample.

Galaxy No. 57 was detected in [OII], [OIII], and H$\beta$.  Though
the signal--to--noise ratio is quite good we can see little extension
beyond the nuclear region of the galaxy.  The width of the emission
lines in the nucleus, on the order of 150 \kms, suggest that this is
a Seyfert class object (Osterbrock 1989).  Dressler and Gunn (1992)
classify this object as having an exponential light distribution. It 
appears roughly spherical in Figure 1 of Dressler and Gunn (1992) 
and lies at some distance from the cluster core.

Galaxy No. 208 is the object used by Dressler and Gunn in their Figure 8
as an example of an emission--line galaxy. We detected both oxygen
lines and H$\beta$, but no emission was detectable beyond the nucleus.
It appears roughly spherical and the light profile follows a r$^{1/4}$
law.

\noindent {\it BOW No. 85 in Abell 963: \ } Our final source is a member of
the rich cluster Abell 963 at $z$ = 0.201, labeled No. 85 in the catalog of
Abell 963 presented by Butcher \etal (1983; hereafter BOW).  We successfully
traced the extent of the [OII] doublet, and detected the fainter H$\beta$ and
[OIII] lines.  The variation in velocity along the major axis drawn from the
[OII] data is presented in Figure 3. We derive a velocity width W of 230
$\pm 8$ \kms.

Lavery and Henry (1988) obtained spectra and images of this blue
emission--line galaxy. Their spectrum shows strong [OII] and [OIII]
emission lines as well as high order Balmer absorption lines. Their
image clearly indicates that the object is elongated, suggesting its
spiral nature.  Although their image is not very deep, they suspect
tidal interaction between this galaxy and a companion about 4\arcsec \
to the north.  Using their image to estimate an axial ratio of 0.44, we
derive a corrected velocity width $W_c$ of 253 $\pm 8$ \kms.

\section{Discussion}

The results of this pilot effort demonstrate the potential of
investigating the kinematics of galaxies at intermediate redshift via
long--slit spectroscopy.  Given very good seeing ($<$ 1\arcsec) and
adequate weather conditions, this project confirms the efficacy of the
measurement of rotation curves in distant galaxies.  The rotation
curves shown in figures 1 and 3 are well-defined and appear to flatten,
characteristics seen in nearby normal spiral galaxies.  At the same
time the presence of strong emission in the objects makes them quite
unusual compared to low redshift galaxies.  Although both galaxies
showing extended emission appear to have nearby companions, there is no
evidence of interaction in their rotation curves.

In order to compare the extent of the rotation curve derived from the various
lines, we obtained similar high resolution long slit spectra for a small
sample of relatively nearby, normal galaxies ($z \sim$ 0.02) over wavelength
ranges covering all of the relevant lines including H$\alpha$. Although the
current sample is still small, primarily because of weather limitations, we
note that the extent of the rotation curve traced in all lines is similar
within the constraint that the signal--to--noise ratio is much higher for the
H$\alpha$ emission.  The velocity widths measured from different lines are
consistent within the errors.  In the nearby objects, the oxygen lines are
far weaker than the H$\alpha$ line, but unlike our high $z$ candidates
these galaxies had not been pre--selected for strong oxygen emission.  

Measurements of the velocity widths allow estimation of galaxy properties
and/or distances via application of the TF relation. Since most TF studies of
nearby galaxies have been calibrated using radio 21--cm line widths, the
optical widths must be transformed to a system appropriate to the radio TF
relation.  This is particularly critical for distant studies where the
velocity width rather than a detailed rotation curve may be all that is
available.  A relative calibration between optical and HI line widths has
been derived from a sample of low redshift galaxies, most of them in nearby
Abell clusters.  Using 145 objects with both high quality HI line widths and
optical velocity widths derived in the same manner employed here, we have
derived a transformation relation between the optical width W and the 21--cm
line width as used by Pierce and Tully (1988).  Using this transformation, we
derive an equivalent corrected (for turbulent broadening) 21--cm line width
$W_c^{21}$ of 185 $\pm$ 15 \kms for Galaxy 2545.3 and 264 $\pm$ 11 \kms for BOW
85. A more complete discussion of the comparison of radio and optical width
measurements will be presented elsewhere (Vogt \etal 1993). 
The corrected 21--cm width derived for 2545.3 is rather small, particularly
relative to the mean expected for a large sample of low redshift
objects (Roberts and Haynes 1993). It is possible, given the uncertainties,
that this object may be less inclined than the current estimate, or
that the rotation curve continues to rise significantly, beyond the
radius at which line emission is detected.

Using the TF relation, the intrinsic brightness of distant galaxies can be
compared, albeit crudely, to those nearby. Using the R-band calibration of
the TF relation by Pierce and Tully (1988), a Hubble constant of 75 km
s$^{-1}$ Mpc$^{-1}$, and the Scd galaxy energy distribution of Pence (1976)
for the K-correction, we estimate that BOW 85 and 2545.3 are 0.9$\pm$0.4 and
3.1$\pm$0.7 magnitudes brighter respectively than predicted from the nearby
galaxy sample.  (Another 0.3 magnitude error should be added in quadrature
with the quoted errors to account for the intrinsic scatter in the TF
relation).   Given the various uncertainties and complications in comparing
the distant galaxies to the nearby ones, this increase in the brightness of
distant galaxies relative to nearby ones of the same velocity width should be
considered preliminary. Similar results have been reported by Franx (1993)
who applied the luminosity--velocity dispersion relation to E+A galaxies
in the cluster A665 at z=0.18.

Because of our galaxy selection process and the potential for galaxy
evolution, use of the TF relation to derive a Hubble constant
is not justified at this time.  In the future a better statistical
sample will allow a comparison of the slope of the TF relation at
high and low redshift and the evaluation of the importance of merging
in the blue galaxy population.  High resolution images such as those
obtained recently with HST by Dressler \etal (1993) will give better
estimates of the galaxy morphologies and inclination angles.  Multiband
application of the TF relation will help to investigate galaxy
evolution since long wavelengths should be progressively less affected
by it. Alternatively, if the effects of galaxy evolution can be
disentangled and the local calibration of the TF relation verified, 
the Hubble constant can be determined.

\acknowledgments
This work has been funded by NSF grants AST9014858 and AST9023450. S.C.
is also supported by a Canadian NSERC Postdoctoral Fellowship.  We
thank Nancy Ellman and David Koo for providing information concerning
galaxy 2545.3 in SA68 and Marco Scodeggio for assistance in obtaining
guide star candidates.

\begin{table*}
\caption{Information on observed galaxies.} \label{tbl-1}
\begin{tabular}{llccccccccc}
\tableline
Field & Galaxy & RA & Dec & $z$ & $R$\tablenotemark{a}  & $W$\tablenotemark{b} & t\tablenotemark{c} & seeing & r\tablenotemark{d} & ref \\
& & h \ \ m \ \ s & \ \ \deg \ \ \ \arcmin \ \ \ \arcsec & & \arcsec & \kms & hr & \arcsec & & \\
\tableline
\tableline
SA68 & 2545.3 & 00 14 36.6 & +17 00 32 & 0.211 & 3.5 & 115 & 2.0 & 1.5  & 18.4\tablenotemark{e} & 1\\
CL0949 & DG 57 & 09 49 42.7 & +44 10 46 & 0.357 & -- & -- & 1.5 & 0.9  & 21.7 & 2\\
CL0949 & DG 208 & 09 49 45.0 & +44 08 14 & 0.380 & -- & 150 & 2.0 & 1.1  & 21.3 & 2\\
A963 & BO 85 & 10 14 11.7 & +39 16 30 & 0.201 & 4.0 & 230 & 2.0 & 0.8  & 19.0 \tablenotemark{f}& 3\\
\tableline
\tablenotetext{a}{Maximum radial extent of detected emission.}
\tablenotetext{b}{Optical velocity width. See text for definition.}
\tablenotetext{c}{Integration time.}
\tablenotetext{d}{Gunn r band magnitude or its equivalent.}
\tablenotetext{e}{Derived from J magnitude of 20.12 from Ellman and Koo and assuming
$<$J--F$>$ = 1.4 and $<$V--R$>$ = 0.5}
\tablenotetext{f}{Derived from F magnitude given by BOW using their conversion from F to
R-band and assuming $<$V--R$>$= 0.5.}
\tablerefs {(1) Koo (1985), (2) Dressler and Gunn (1992), (3) Butcher,
Oemler and Wells (1983).}
\end{tabular}
\end{table*}

\clearpage
\begin{figure}
\caption{The spatial variation in the velocities in the rest frame of
the galaxy derived from the [OII] and [OIII] line profiles for 2545.3 in 
the SA 68 field.}
\end{figure}
   
\begin{figure}
\caption{Spectral image of the resolved [OII] doublet ($\lambda$3726, 
$\lambda$3729) for the galaxy 2545.3 in the SA 68 field.  The grid step is 
is 0.56 \AA \ per pixel in the dispersion direction and \hskip 1.0 truein 
0.78 \arcsec \ per pixel in the cross--dispersion direction.  The peak 
intensity is 45--$\sigma$ above the noise.}
\end{figure}

\begin{figure}
\caption{The spatial variation in the velocities in the rest frame of the
galaxy derived from the [OII] line profiles for the No. 85 in Abell 963.
The data at -3\arcsec \ fell onto a bad column on the chip and was
distorted.}
\end{figure}
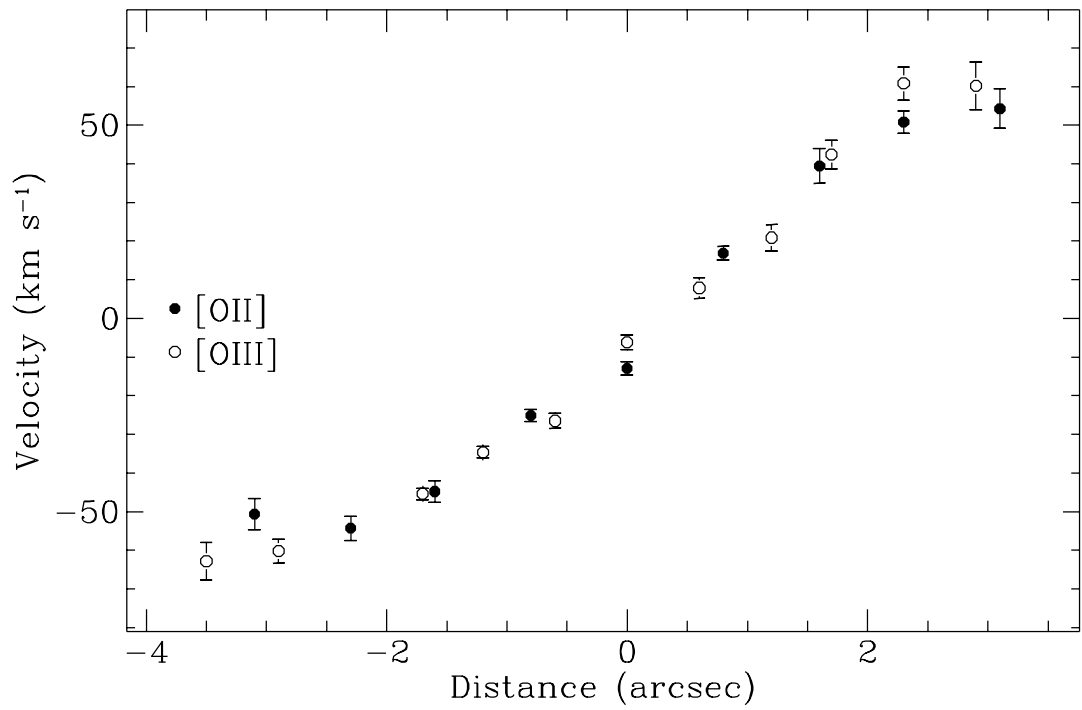
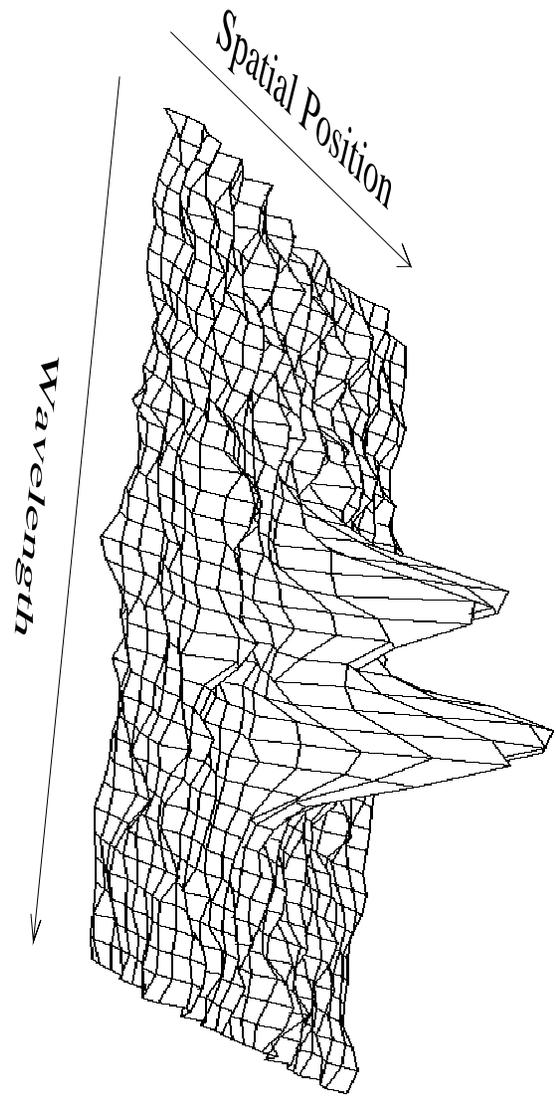
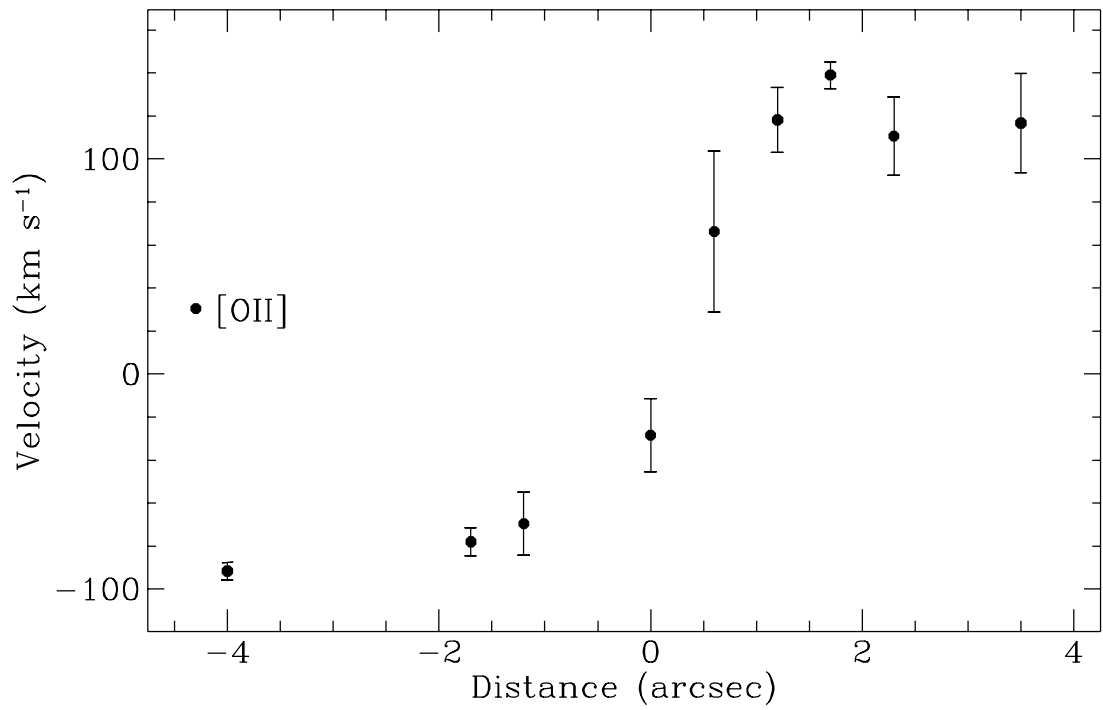
\end{document}